\documentclass[%
 reprint,
 amsmath,amssymb,
 aps,
]{revtex4-1}

\usepackage{graphicx}
\usepackage{dcolumn}
\usepackage{bm}

\begin{document}

\preprint{APS/123-QED}

\title{Broadband Nonreciprocal Amplification in Luminal Metamaterials}

\author{E. Galiffi$^1$}
\author{P. A. Huidobro$^{1,2}$}%
\author{J. B. Pendry$^1$}
\affiliation{$^1$ Condensed Matter Theory Group, The Blackett Laboratory, Imperial College London, UK\\$^2$ Instituto de Telecomunica\c c\~oes, Instituto Superior Tecnico-University of Lisbon, Portugal}




\date{\today}

\begin{abstract}
    Time has emerged as a new degree of freedom for metamaterials, promising new pathways in wave control. However, electromagnetism suffers from limitations in the modulation speed of material parameters. Here we argue that these limitations can be circumvented by introducing a traveling-wave modulation, with the same phase velocity of the waves. We show how luminal metamaterials generalize the parametric oscillator concept, realizing giant, broadband nonreciprocity, achieving efficient one-way amplification, pulse compression and harmonic generation, and propose a realistic implementation in double-layer graphene.
\end{abstract}

\maketitle



Temporal control of light is a long-standing dream, which has recently demonstrated its potential to revolutionize optical and microwave technology, as well as our understanding of electromagnetic theory, overcoming the stringent constraint of energy conservation~\cite{shaltout2019spatiotemporal}. Along with the ability of time-dependent systems to violate electromagnetic reciprocity~\cite{sounas2017non,hadad2016breaking,PhysRevApplied.10.047001}, realising photonic isolators and circulators~\cite{PhysRevLett.110.093901, sounas2013giant,yu2009complete, PhysRevLett.108.153901}, amplify signals \cite{koutserimpas2018parametric}, perform harmonic generation~\cite{chamanara2018linear,deck2018scattering,ginis2015tunable} and phase modulation~\cite{sherrott2017experimental}, new concepts from topological~\cite{lin2016photonic,fleury2016floquet,he2019floquet} and non-Hermitian physics~\cite{koutserimpas2018nonreciprocal,regensburger2012parity} are steadily permeating this field. However, current limitations to the possibility of significantly fast modulation in optics has constrained the concept of time-dependent electromagnetics to the radio frequency domain, where varactors can be used to modulate capacitance~\cite{barnes1961voltage}, and traveling-wave tubes are commonly used as (bulky) microwave amplifiers~\cite{pierce1950traveling}. In the visible and near IR, optical nonlinearities have often been exploited to generate harmonics, and realize certain nonreciprocal effects~\cite{sounas2018broadband}. However, nonlinearity is an inherently weak effect, and high field intensities are typically required.

In this Letter, we challenge the very need for high modulation frequencies, demonstrating that strong and broadband nonreciprocal response can be obtained by complementing the temporal periodic modulation of an electromagnetic medium with a spatial one, in such a way that the resulting traveling-wave modulation profile appears to drift uniformly at the speed of the wave, i.e., a \textit{luminal} modulation.
We show that unidirectional amplification and compression can be accomplished in luminal metamaterials, which thus constitute a broadband generalization of the narrowband concept of the parametric oscillator, enabling harmonic generation with exponential efficiency. We present a realistic implementation based on acoustic plasmons in double-layer graphene (DLG), thus circumventing the intrinsic limitations in the modulation speed of its doping level. Our findings, which are transferable to other wave domains, hold potential for efficient harmonic generation (terahertz, in the specific case of graphene), loss-compensation and amplification of waves. \begin{figure}[b!]
    \begin{center}
        \includegraphics[width=0.7\columnwidth]{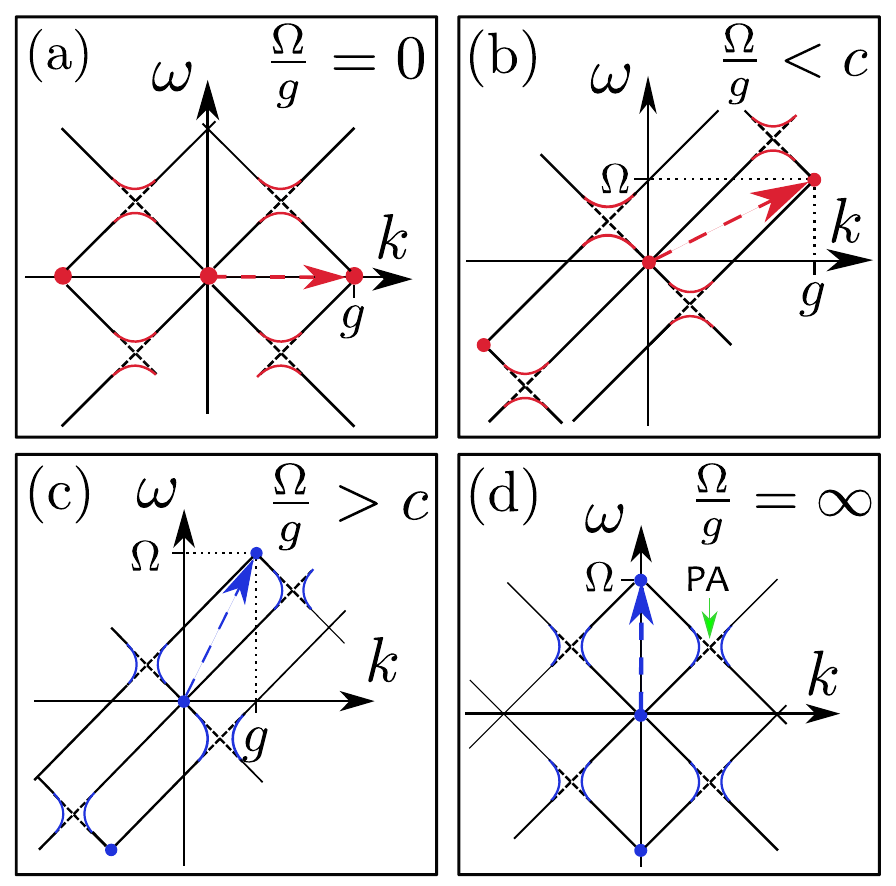}
        \caption{\label{fig:band_structure} (a) The band structure of a conventional spatial crystal is repeated in phase space at $k = n g$, $n\in \mathbb{Z}$, forming vertical band gaps ($\omega$-gaps). (b) Similarly, the band structure of a traveling-wave-modulated crystal is symmetric under discrete translations by an oblique reciprocal lattice vector $(g,\Omega)$. When $\Omega/g<c$, $\omega$-gaps open, whereas (c) $\Omega/g>c$ leads to unstable $k$-gaps. (d) Finally, if the wavelength of the modulation $L\to \infty$, then $g\to 0$, so that the system is effectively only modulated in time  In this case, the modulation speed $\frac{\Omega}{g}\to \infty$ and the system becomes a narrowband, reciprocal, parametric amplifier. The transition between (b) and (c), whereby the light-line and the reciprocal lattice vector are aligned, is a luminal crystal.}
    \end{center}
\end{figure} 

Bloch (Floquet) theory dictates that the wavevector (frequency) of a monochromatic wave propagating in a spatially (temporally) periodic medium can only Bragg-scatter onto a discrete set of harmonics, determined by the reciprocal lattice vectors. This still holds true when the modulation is of a travelling-wave type, whereby Bragg scattering couples Fourier modes which differ by a discrete amount of both energy and momentum~\cite{cassedy1963dispersion,cassedy1967dispersion,biancalana2007dynamics,yu2009complete,sounas2017non,PhysRevB.96.165144}. \begin{figure*}[t!]
    \begin{center}
        \includegraphics[width=1.7\columnwidth]{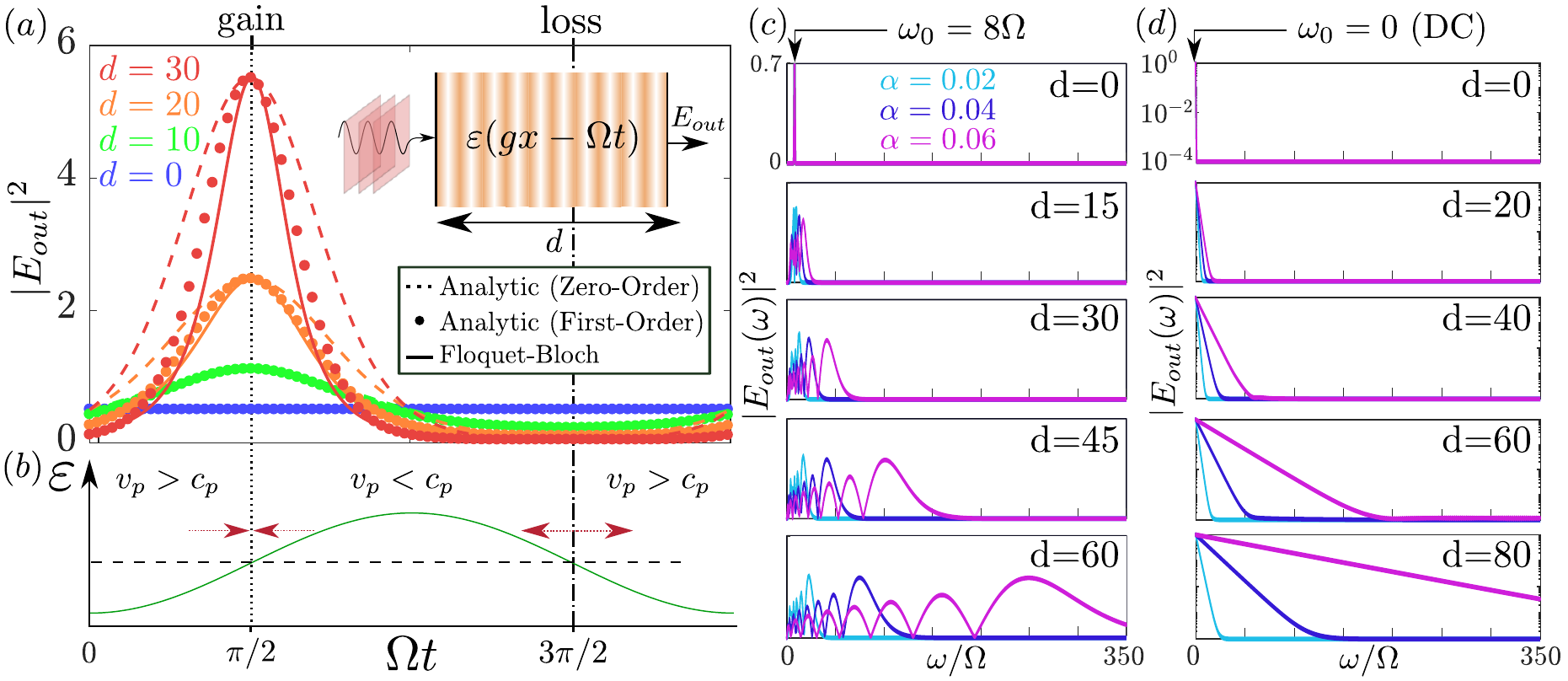}
        \caption{\label{fig:amplification} (a) An incident plane wave is concentrated and exponentially amplified as it propagates through a luminal metamaterial ($g=\Omega=1$, $\alpha=0.04$) of length $d$ (inset), at whose exit the field is calculated. Continuous lines correspond to Floquet-Bloch theory, whereas dashed lines and circles were obtained from our analytic model to zeroth and first (Eq. \ref{eq:intensity}) order respectively. (b) Waves preceding (following) the gain-point ($\Omega t = \pi/2$) experience a lower (higher) permittivity, hence a higher (lower) phase velocity, thus being attracted towards the gain-point, at which amplification occurs. Conversely, waves preceding (following) the loss-point ($\Omega t = 3\pi/2$) are drawn away from it, depleting it of energy. (c) An incident monochromatic wave with input frequency $\omega_0=8\Omega$ is efficiently coupled to higher harmonics at an exponential rate. Beating arises from the different $\Omega$ and $\omega_0$ (d) The frequency content (log-scale) of a DC input applied to a luminal metamaterial spreads out exponentially in Fourier space, generating a supercontinuum.}
    \end{center}
\end{figure*} As shown in Fig.~\ref{fig:band_structure} for a 1D system, these space-time reciprocal lattice vectors can be defined to take any angle in phase space, depending on whether a generic traveling-wave modulation of the material parameters of the form $\delta \epsilon(gx-\Omega t)$ is spatial (panel a: $\Omega=0$ ), temporal  (d: $g=0$), or spatiotemporal (b,c: $g\neq 0$, $\Omega\neq 0$). Given the slope $c$ of the bands in a Brillouin diagram, which denotes the velocity of waves in a dispersionless medium, the speed of the traveling-wave modulation defines a subluminal regime $\frac{\Omega}{g}<c$ (a,b), whereby conventional vertical band gaps open~\cite{cassedy1963dispersion}, and a superluminal one $\frac{\Omega}{g}>c$ (c,d), characterized by horizontal, unstable $k$-gaps~\cite{cassedy1967dispersion,PhysRevA.57.4905}. A common example of the latter is the parametric amplifier ($g=0$: Fig.~\ref{fig:band_structure}d): when the parameters governing an oscillatory system are periodically driven at twice its natural frequency, exponential amplification occurs, as a result of the unstable $k$-gap at frequency $\omega = \Omega/2$. However, achieving such fast modulation at infrared frequencies remains a key challenge for dynamical metamaterials.

The transition between the regimes in Figs.~\ref{fig:band_structure}b and \ref{fig:band_structure}c, i.e. $\Omega/g=c$, is an exotic degenerate state that we name luminal metamaterial, whereby all forward-propagating modes are uniformly coupled. Due to its broadband spectral degeneracy in the absence of dispersion, this system is highly unstable, thus preventing a meaningful definition of its band structure. Nevertheless, if we consider transmission through a spatially (temporally) finite system with well-defined boundary conditions, causality can be imposed in the unmodulated regions of space (time), so that an expansion into eigenfunctions can be performed, as detailed in the S.M.~\cite{supp_mat}. In luminal metamaterials, the photonic transitions induced by the modulation of the refractive index are no longer interband~\cite{winn1999}, but intraband, and can therefore be driven by means of any refractive index modulation, regardless of how adiabatic, whose reciprocal lattice vector $(g,\Omega)$ satisfies the speed-matching condition $\Omega/g=c$. Hence, any limitation in modulation frequency $\Omega$ can be compensated, in principle, by a longer spatial period $L=2\pi/g$. Notably, these can be locally induced by modulating the properties of the medium, and can thus synthetically move at any speed, including and exceeding the speed of light, in analogy with the touching point of a water wave front propagating almost perpendicularly to a beach, or the junction between the blades of a pair of scissors. 
 
In real space, amplification in this system can be modelled as follows: consider a non-dispersive, lossless medium where $\varepsilon(x,t) = 1+2\alpha\cos(gx-\Omega t)$, with $\Omega/g = c_0$. Following the derivation of Poynting's theorem, we can write: \begin{align}
    \nabla \cdot (\mathbf{E} \times \mathbf{H})  &= -\frac{\mu_0}{2} \frac{\partial H^2}{\partial t} - \frac{\varepsilon_0\varepsilon}{2}\frac{\partial E^2}{\partial t} - \varepsilon_0 \frac{\partial \varepsilon}{\partial t} E^2,
\end{align} so that the total time-derivative of the local energy density is: \begin{align}
    \frac{d U}{d t} = -\frac{1}{\varepsilon} \frac{\partial \varepsilon}{\partial t} U- \frac{\partial P}{\partial x} + c_0 \frac{\partial U}{\partial x} = -\frac{1}{\varepsilon} \frac{\partial \varepsilon}{\partial t} U - \frac{\partial P'}{\partial x},
    \label{eq:en_dens_equation}
\end{align} where the compensated Poynting vector, $P'$, consists of a local and an advective part (due to the moving frame)~\cite{supp_mat}. The first term in Eq.~\ref{eq:en_dens_equation} is responsible for gain, whereas the second describes the Poynting flux, which drives the compression of the pulse. Ignoring the Poynting contribution to zero-order yields $U(X,t) = e^{-2 \alpha \Omega t \sin(gX)}$, where $X = x-\Omega t/g$. Feeding the zero-order solution into the resulting compensated Poynting vector $P' = c_0(\varepsilon(X,t)^{-\frac{1}{2}}-1) U $ in Eq.~\ref{eq:en_dens_equation}, we obtain a corrected expression for the energy density:
\begin{align}
    U(X,t) = \exp{[-2\alpha \Omega t \sin(gX) - \alpha^2 \Omega^2 t^2 \cos^2(gX)]}. \label{eq:intensity}
\end{align}{} Alternatively, the system can also be modelled with a semi-analytic Floquet-Bloch expansion of the fields, and the transmission coefficient can be calculated for a finite slab, validating our analytical expressions~\cite{supp_mat}. Assuming a slab of length $d$, and substituting $\Omega t = g d$ in Eq.~\ref{eq:intensity}, we calculate the temporal profile of the electric field intensity at the output $x = d$ (Fig.~\ref{fig:amplification}a). The modulation is able to exponentially amplify and concentrate the signal at the point with phase $\Omega t = \pi/2$, and exponentially suppress it at $\Omega t = 3\pi/2$. The reason is apparent from Fig.~\ref{fig:amplification}b: those field amplitudes which sit at $-\pi/2 < \Omega t < \pi/2$ experience a lower permittivity, and hence a higher phase velocity, whereas those sitting at $\pi/2<\Omega t<3\pi/2$ lag, so that the point corresponding to a phase $\Omega t = \pi/2$ acts as an attractor, or gain-point, where the modulation imparts energy into the wave. Conversely, $\Omega t = 3\pi/2$ is a repellor, or loss-point, where energy is absorbed by the modulation drive (further numerical simulations are provided in the S.M.~\cite{supp_mat}).

As evidenced by the absence of any frequency dependence in Eq.~\ref{eq:intensity}, and in contrast to conventional time-modulated systems, parametric amplification in a luminal medium is a fully broadband phenomenon, enabling exponentially efficient generation of frequency-wavevector harmonics, as shown in Fig.~\ref{fig:amplification}c. Remarkably, even a DC input can be transformed into a broadband pulse train at an exponential rate, as revealed by Floquet-Bloch calculations (see Fig.~\ref{fig:amplification}d). Our closed-form analytic solution enables us to exactly quantify the power amplification rate as $2\alpha\Omega$, which needs to overcome the loss, for amplification to occur. However, the reactive behaviour responsible for the compression performance is unaffected by losses, which only reduce the overall output power efficiency. Furthermore, these systems are transparent to counterpropagating waves, thus entailing the additional advantage of nonreciprocity. Moreover, while nonreciprocal response is typically observed only near band gaps in conventional systems~\cite{sounas2017non}, it is achieved at virtually any frequency in a luminal metamaterial. 

Due to their ease of manipulation, metasurfaces offer the most promising playground to realize dynamical effects~\cite{shaltout2019spatiotemporal,correas2015nonreciprocal,dinc2017synchronized}, also due to the rise of tunable two-dimensional materials~\cite{novoselov20162d,basov2016polaritons}.
Recently, graphene has emerged as a platform to enhance light-matter interactions~\cite{koppens2011graphene,PhysRevB.84.195446,grigorenko2012graphene,ju2011graphene}, realizing atomically thin metasurfaces~\cite{vakil2011transformation,slipchenko2013analytical,huidobro2016graphene,poumirol2017electrically,sherrott2017experimental}. Its doping level, which can be tuned with ion-gel techniques to be as high as $2$ eV~\cite{chen2011controlling,efetov2010controlling}, can be dynamically modulated via all-optical techniques, with experimentally reported response times as short as $2.2$ ps at relative doping modulation amplitudes of $38\%$~\cite{li2014ultrafast,tasolamprou2019experimental}. In addition, modern-quality graphene features extremely high electron mobility, with measured experimental values of $350'000$ cm$^2$/(V$\cdot$s)~\cite{banszerus2015ultrahigh}.  

\begin{figure}[t!]
    \begin{center}
        \includegraphics[width=\columnwidth]{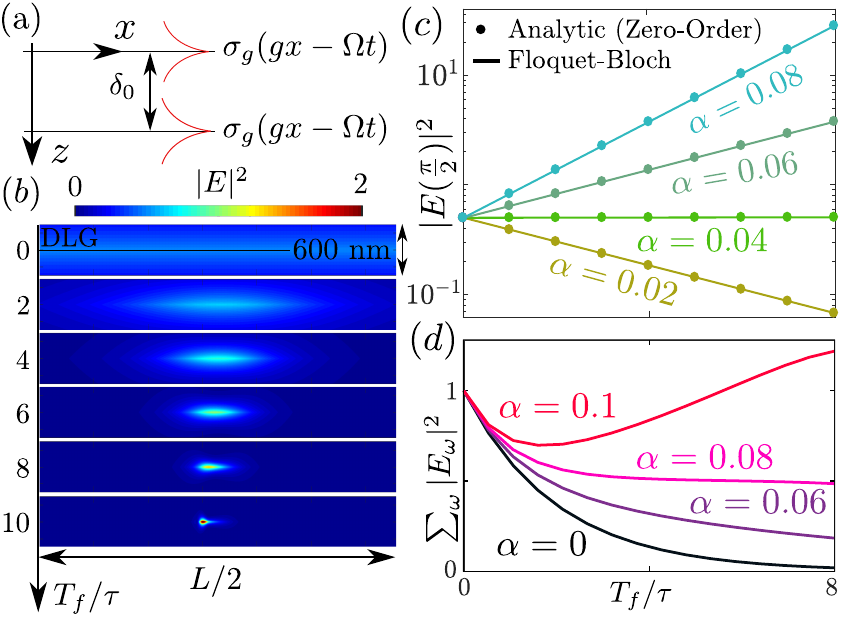}
        \caption{\label{fig:graphene}(a) Double-layer graphene configuration. (b) The DLG plasmon intensity is amplified and compressed at the gain-point $gX = \pi/2$, as a luminal modulation is applied over increasingly long time-windows $T_f$. (c) The intensity at the gain-point grows exponentially as a function of both modulation time $T_f$ and amplitude $\alpha$ for sufficiently strong  ($\alpha = 0.06, 0.08$) or fast modulation, as predicted by our analytic model. (d) The integrated power of the plasmon reduces initially, due to dissipation. Once the pulse is localised near the gain-point, loss compensation ($\alpha = 0.06, 0.08$), and even amplification ($\alpha = 0.1$) are possible.}
    \end{center}
\end{figure} The dispersion relation of graphene plasmons follows a square root behaviour $\omega\sim \sqrt{k}$, where $\omega$ is the angular frequency and $k$ the in-plane wavevector. However, in a double-layer configuration (Fig.~\ref{fig:graphene}a), a second, ``acoustic" plasmon branch arises~\cite{gonccalves2016introduction,iranzo2018probing,alonso2017acoustic}, whose dispersion:\begin{align}
    \omega \propto \sqrt{\epsilon_F} \sqrt{k(1-e^{-\delta_0 k})} \simeq  \sqrt{\delta_0\epsilon_F}k(1-\frac{\delta_0 k}{4})
\end{align} is linear for small interlayer gaps $\delta_0 \ll k^{-1}$ ($\epsilon_F$ is the Fermi energy). Here we exploit the linearity of this acoustic plasmon band to realize a luminal metasurface,  whilst accounting for dispersion, and we demonstrate nonreciprocal plasmon amplification and compression. Alternative amplification schemes for graphene plasmons have been theoretically proposed, such as drift currents~\cite{morgado2017negative,PhysRevB.97.085419,ghafarian2016millimetre}, periodic doping modulation~\cite{wilson_low2018}, adiabatic doping suppression~\cite{sun2016adiabatic}, and plasmonic \v{C}erenkov emission by hot carriers~\cite{kaminer2016efficient}.

We assume a semiclassical (Drude) conductivity model, which is accurate as long as $\hbar \omega \ll \epsilon_F$ and $k\ll k_F$. Our setup consists of two graphene layers, whose Fermi levels are modulated as $\epsilon_F(x,t) = \epsilon_{F,0}[1 + 2\alpha \cos(gx-\Omega t)]$ (Fig.~\ref{fig:graphene}a). Dispersion is accounted for, by expressing the constitutive relation for the current $J(x,t)$ in Fourier space, where the conductivity modulation couples neighbouring frequency harmonics:
\begin{align}
J_n &= \frac{e^2 \epsilon_{F,0}}{\pi\hbar^2}  \frac{E_n+\alpha (E_{n+1} + E_{n-1})}{\gamma-i(\omega+n\Omega)}, \label{eq:current_electric}
\end{align} 
where $\gamma$ is the loss rate and $E_n$ is the n$^{th}$ Fourier amplitude of the in-plane electric field, which is continuous at the layer positions $z=0$ and $z=\delta_0$, as detailed in the S.M.~\cite{supp_mat}. The magnetic field of the p-polarized wave $H_y(x,z,t)$ is discontinuous at the layers by the surface current~\cite{gonccalves2016introduction}. This system can be accurately described within an adiabatic regime, since the modulation frequency $\Omega \ll \omega$. Furthermore, since acoustic plasmons carry much larger momentum than photons, the modes are strongly quasistatic, so that the out of plane decay constant $\kappa_n\simeq k+ng$, and coupling to radiation is negligible, given that both spatial and temporal frequencies of the doping modulation are much smaller than the plasmon wavevector and frequency. Taking advantage of the adiabatic assumption, we can conveniently solve the scattering problem in the time-domain, as detailed in the S.M.~\cite{supp_mat}.

In our calculations we assume a Fermi energy $\epsilon_F=1.5$ eV $ \approx 2 \pi\hbar \times 362$ THz, and a loss rate $\gamma=\frac{v_F^2 e}{m\epsilon_{F,0}}\approx 60$ GHz, where $m=10^{5}$ cm$^2$/(V$\cdot$s) is the electron mobility, and the Fermi velocity $v_F\approx 9.5 \times 10^5$ms$^{-1}$. Fig.~\ref{fig:graphene}b demonstrates plasmon amplification and compression for different modulation times $T_f$. Here, we use a modulation amplitude $\alpha = 0.05$, interlayer gap $\delta_0 = 1$ nm, an input frequency $\omega/2\pi = 1$ THz and a modulation frequency $\Omega/2\pi = 120$ GHz, which corresponds to a modulation period $\tau = 2\pi/\Omega \approx 8$ ps and length $L \approx 26$ $\mu$m, such that the long-wavelength phase velocity of the plasmon is matched by the modulation speed $c_p = \Omega/g$. Since the DLG plasmon bands are approximately linear, we can set $c_0 = c_p$ in our closed-form solution (Eq.~\ref{eq:intensity}), and verify the analogous amplification mechanism, showing excellent agreement  with Floquet-Bloch theory (Fig.~\ref{fig:graphene}c). Finally, Fig.~\ref{fig:graphene}d demonstrates the total power amplification achieved by our luminal graphene metasurface: initially the unit input power of the wave is predominantly dissipated by the uniform losses, except at the gain-point, so that this first propagation moment is dominated by damping. Once sufficient power is accumulated at the gain-point, the energy fed by the modulation into the plasmon ensures that its propagation is effectively loss-compensated, as in the case of $\alpha=0.08$, extending its lifetime by orders of magnitude, or even amplifying it, as in the $\alpha = 0.1$ case.

As the luminal modulation couples the frequency content of the pulse to very high frequency-wavevector harmonics, these will experience the nonlinearity of the bands.\begin{figure}[t!]
    \begin{center}
        \includegraphics[width=\columnwidth]{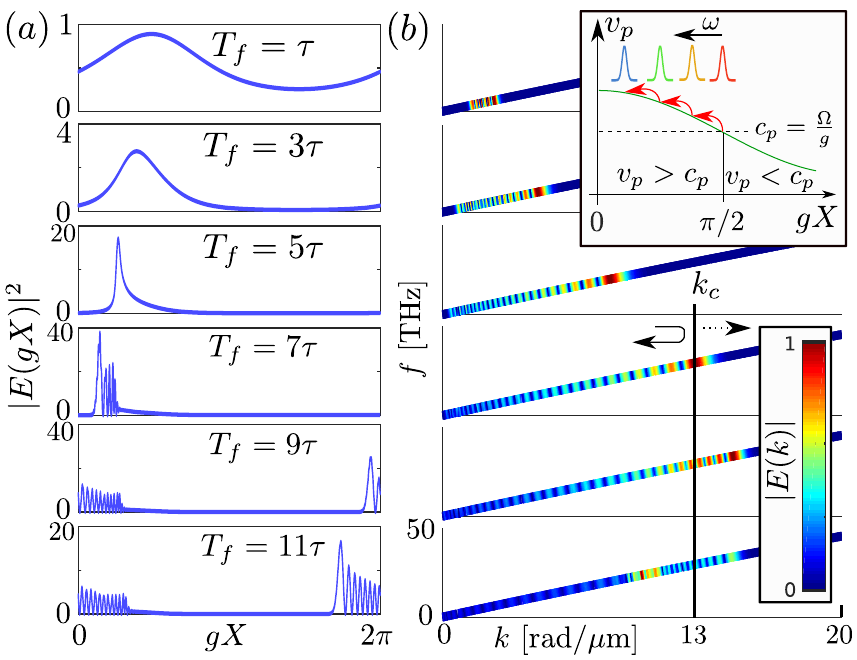}
        \caption{\label{fig:dispersion} (a) The effect of dispersion for a wider gap $\delta_0 = 15$ nm causes a gradual shift of the gain-point due to the slower phase velocity of higher frequency components, as well as a skewing on the pulse. For even longer modulation times, the wave breaks into a train of narrow pulses. (b) The spectral content of the pulse is amplified and projected from an input frequency of $\approx 4$ THz, to $\approx 30$ THz, demonstrating efficient terahertz frequency generation using a modulation frequency of only $\Omega/2\pi = 120$ GHz. High frequency components whose phase velocity is below the instability threshold are not coupled to, hence dispersion stabilizes the system. Here we assumed a wider gap $\delta_0 = 15$ nm to highlight the effect of dispersion, $\alpha = 0.05$.}
    \end{center}
\end{figure} In Fig.~\ref{fig:dispersion} we use a wider inter-layer gap $\delta_0 = 15$ nm and higher mobility $m=10^{6}$ cm$^{2}$/(V$\cdot$s), to highlight the effects of dispersion on the pulse profile (a) and its spectral content (b) for different modulation times $T_f$. At a first stage, since higher frequency components experience a slightly lower phase velocity, the gain-point must shift back to $gX<\frac{\pi}{2}$, where the increase in local phase velocity determined by the modulated Fermi energy compensates for the curvature of the band (Fig.~\ref{fig:dispersion}, inset). In addition, Fourier components propagating with phase velocity $c_p$ are amplified near the conventional gain-point, thus skewing the pulse ($T_f = 5\tau$). Finally, for even longer propagation times, the wave will cease to compress, and break into a train of pulses. This is due to the existence of a finite regime of phase velocities: $(1+2\alpha)^{-1/2} < v_p/c_p < (1-2\alpha)^{-1/2}$, within which the interaction between co-propagating bands is strong enough to make the system unstable~\cite{cassedy1967dispersion}. In our setup, the relative phase velocity, \begin{align}
    \frac{v_p(k)}{c_p} =\big(\frac{\omega(k)}{k}\big)\big/\big(\frac{\Omega}{g}\big)  \simeq (\frac{1-e^{-k\delta_0}}{\delta_0 k})^{1/2},
\end{align} decreases approximately linearly with increasing wavevector~\cite{supp_mat}. Equating the latter to the lower threshold velocity ratio $v_p(k_c)/c_p = \frac{1}{\sqrt{1+2\alpha}}$, where $\alpha = 0.05$ and expanding the exponential to second order, we get an analytical estimate for the critical wavevector $k_c\approx 13$ rad$/\mu$m, beyond which the pulse is no longer strongly coupled to higher harmonics, and its power spectrum is effectively reflected, resulting in beating. Thus, dispersion plays the important role of stabilizing these systems. Subsequently, the power spectrum oscillates within the extended luminal region, although beating between different space-time harmonics, no longer in-phase, induces fast oscillations, reminiscent of comb formation in nonlinear optics~\cite{knight1992optical}. 

In this Letter, we have introduced the concept of luminal metamaterials, realized by inducing a traveling-wave modulation in the permittivity of a material, whose phase velocity matches that of the waves propagating in it, in the absence of modulation. We have shown that these dynamical structures generalize the concept of parametric amplification to cover a virtually unlimited bandwidth, thus being capable of reinforcing and compressing input waves of any frequency, including a DC field. We have demonstrated their robustness against moderate dispersion, and proposed a realistic implementation exploiting acoustic plasmons in double-layer graphene, thus paving a new viable route towards the amplification of graphene plasmons, and terahertz generation. Furthermore, luminal metamaterials exhibit an inherent, strongly nonreciprocal response at any frequency, due to the directional bias induced by the modulation, whose phase velocity can be made as high as needed by extending the spatial period of the modulation, the only limitation being the propagation length of the excitation, and hence the loss. 

Furthermore, thanks to its ability to couple incident electromagnetic waves to higher frequency-momentum harmonics at an exponential rate, the luminal metamaterial concept constitutes a fundamentally new path towards efficient harmonic generation, which can work even with a DC input, necessitating only of low modulation speeds, as opposed to conventional parametric systems. Finally, we remark that this concept can be translated to any wave system which exhibits a linear or weakly dispersive regime, such as acoustic, elastic and shallow-water waves, and the reach of this mechanism could be further extended by introducing chirping, in analogy with the tuning of the frequency of a driving field with the energy of electrons accelerated in a synchrocyclotron.

\begin{acknowledgments}
E.G. acknowledges support through a studentship in the Centre for Doctoral Training on Theory and Simulation of Materials at Imperial College London funded by the EPSRC (EP/L015579/1). P.A.H. acknowledges funding from Funda\c c\~ao para a Ci\^encia e a Tecnologia and Instituto de Telecomunica\c c\~oes under projects CEECIND/03866/2017 and UID/EEA/50008/2019. J.B.P. acknowledges funding from the Gordon and Betty Moore Foundation.
\end{acknowledgments}

\bibliographystyle{unsrt} 
\bibliography{LuminalGrapheneBiblio}

\providecommand{\noopsort}[1]{}\providecommand{\singleletter}[1]{#1}%
\begin{thebibliography}{10}

\bibitem{shaltout2019spatiotemporal}
Amr~M Shaltout, Vladimir~M Shalaev, and Mark~L Brongersma.
\newblock Spatiotemporal light control with active metasurfaces.
\newblock {\em Science}, 364(6441):eaat3100, 2019.

\bibitem{sounas2017non}
Dimitrios~L Sounas and Andrea Al{\`u}.
\newblock Non-reciprocal photonics based on time modulation.
\newblock {\em Nature Photonics}, 11(12):774, 2017.

\bibitem{hadad2016breaking}
Yakir Hadad, Jason~C Soric, and Andrea Alu.
\newblock Breaking temporal symmetries for emission and absorption.
\newblock {\em Proceedings of the National Academy of Sciences},
  113(13):3471--3475, 2016.

\bibitem{PhysRevApplied.10.047001}
Christophe Caloz, Andrea Al\`u, Sergei Tretyakov, Dimitrios Sounas, Karim
  Achouri, and Zo\'e-Lise Deck-L\'eger.
\newblock Electromagnetic nonreciprocity.
\newblock {\em Phys. Rev. Applied}, 10:047001, Oct 2018.

\bibitem{PhysRevLett.110.093901}
Da-Wei Wang, Hai-Tao Zhou, Miao-Jun Guo, Jun-Xiang Zhang, J\"org Evers, and
  Shi-Yao Zhu.
\newblock Optical diode made from a moving photonic crystal.
\newblock {\em Phys. Rev. Lett.}, 110:093901, Feb 2013.

\bibitem{sounas2013giant}
Dimitrios~L Sounas, Christophe Caloz, and Andrea Alu.
\newblock Giant non-reciprocity at the subwavelength scale using angular
  momentum-biased metamaterials.
\newblock {\em Nature Communications}, 4:2407, 2013.

\bibitem{yu2009complete}
Zongfu Yu and Shanhui Fan.
\newblock Complete optical isolation created by indirect interband photonic
  transitions.
\newblock {\em Nature Photonics}, 3(2):91, 2009.

\bibitem{PhysRevLett.108.153901}
Kejie Fang, Zongfu Yu, and Shanhui Fan.
\newblock Photonic aharonov-bohm effect based on dynamic modulation.
\newblock {\em Phys. Rev. Lett.}, 108:153901, Apr 2012.

\bibitem{koutserimpas2018parametric}
Theodoros~T Koutserimpas, Andrea Al{\`u}, and Romain Fleury.
\newblock Parametric amplification and bidirectional invisibility in
  pt-symmetric time-floquet systems.
\newblock {\em Physical Review A}, 97(1):013839, 2018.

\bibitem{chamanara2018linear}
Nima Chamanara and Christophe Caloz.
\newblock Linear pulse compansion using co-propagating space-time modulation.
\newblock {\em arXiv preprint arXiv:1810.04129}, 2018.

\bibitem{deck2018scattering}
Zo{\'e}-Lise Deck-L{\'e}ger, Nima Chamanara, Maksim Skorobogatiy, M{\'a}rio~G
  Silveirinha, and Christophe Caloz.
\newblock Scattering in subluminal and superluminal space-time crystals.
\newblock {\em arXiv preprint arXiv:1808.02863}, 2018.

\bibitem{ginis2015tunable}
Vincent Ginis, Philippe Tassin, Thomas Koschny, and Costas~M Soukoulis.
\newblock Tunable terahertz frequency comb generation using time-dependent
  graphene sheets.
\newblock {\em Physical Review B}, 91(16):161403, 2015.

\bibitem{sherrott2017experimental}
Michelle~C Sherrott, Philip~WC Hon, Katherine~T Fountaine, Juan~C Garcia,
  Samuel~M Ponti, Victor~W Brar, Luke~A Sweatlock, and Harry~A Atwater.
\newblock Experimental demonstration of> 230 phase modulation in gate-tunable
  graphene--gold reconfigurable mid-infrared metasurfaces.
\newblock {\em Nano letters}, 17(5):3027--3034, 2017.

\bibitem{lin2016photonic}
Qian Lin, Meng Xiao, Luqi Yuan, and Shanhui Fan.
\newblock Photonic weyl point in a two-dimensional resonator lattice with a
  synthetic frequency dimension.
\newblock {\em Nature Communications}, 7:13731, 2016.

\bibitem{fleury2016floquet}
Romain Fleury, Alexander~B Khanikaev, and Andrea Alu.
\newblock Floquet topological insulators for sound.
\newblock {\em Nature Communications}, 7:11744, 2016.

\bibitem{he2019floquet}
Li~He, Zachariah Addison, Jicheng Jin, Eugene~J Mele, Steven~G Johnson, and
  Bo~Zhen.
\newblock Floquet chern insulators of light.
\newblock {\em arXiv preprint arXiv:1902.08560}, 2019.

\bibitem{koutserimpas2018nonreciprocal}
Theodoros~T Koutserimpas and Romain Fleury.
\newblock Nonreciprocal gain in non-hermitian time-floquet systems.
\newblock {\em Physical Review Letters}, 120(8):087401, 2018.

\bibitem{regensburger2012parity}
Alois Regensburger, Christoph Bersch, Mohammad-Ali Miri, Georgy Onishchukov,
  Demetrios~N Christodoulides, and Ulf Peschel.
\newblock Parity--time synthetic photonic lattices.
\newblock {\em Nature}, 488(7410):167, 2012.

\bibitem{barnes1961voltage}
Sanford~H Barnes and John~E Mann.
\newblock Voltage sensitive semiconductor capacitor, June~20 1961.
\newblock US Patent 2,989,671.

\bibitem{pierce1950traveling}
John~Robinson Pierce.
\newblock Traveling-wave tubes.
\newblock {\em The bell System technical journal}, 29(2):189--250, 1950.

\bibitem{sounas2018broadband}
Dimitrios~L Sounas, Jason Soric, and Andrea Alu.
\newblock Broadband passive isolators based on coupled nonlinear resonances.
\newblock {\em Nature Electronics}, 1(2):113, 2018.

\bibitem{cassedy1963dispersion}
ES~Cassedy and AA~Oliner.
\newblock Dispersion relations in time-space periodic media: Part i—stable
  interactions.
\newblock {\em Proceedings of the IEEE}, 51(10):1342--1359, 1963.

\bibitem{cassedy1967dispersion}
ES~Cassedy.
\newblock Dispersion relations in time-space periodic media part ii—unstable
  interactions.
\newblock {\em Proceedings of the IEEE}, 55(7):1154--1168, 1967.

\bibitem{biancalana2007dynamics}
Fabio Biancalana, Andreas Amann, Alexander~V Uskov, and Eoin~P O’reilly.
\newblock Dynamics of light propagation in spatiotemporal dielectric
  structures.
\newblock {\em Physical Review E}, 75(4):046607, 2007.

\bibitem{PhysRevB.96.165144}
Sajjad Taravati, Nima Chamanara, and Christophe Caloz.
\newblock Nonreciprocal electromagnetic scattering from a periodically
  space-time modulated slab and application to a quasisonic isolator.
\newblock {\em Phys. Rev. B}, 96:165144, Oct 2017.

\bibitem{PhysRevA.57.4905}
M.~Blaauboer, A.~G. Kofman, A.~E. Kozhekin, G.~Kurizki, D.~Lenstra, and
  A.~Lodder.
\newblock Superluminal optical phase conjugation: Pulse reshaping and
  instability.
\newblock {\em Phys. Rev. A}, 57:4905--4912, Jun 1998.

\bibitem{supp_mat}
See supplemental material, which includes
  refs.~\cite{cassedy1965waves,stauber2007electronic,gonccalves2016introduction}.

\bibitem{winn1999}
Joshua~N. Winn, Shanhui Fan, John~D. Joannopoulos, and Erich~P. Ippen.
\newblock Interband transitions in photonic crystals.
\newblock {\em Phys. Rev. B}, 59:1551--1554, Jan 1999.

\bibitem{correas2015nonreciprocal}
D~Correas-Serrano, JS~Gomez-Diaz, DL~Sounas, Y~Hadad, A~Alvarez-Melcon, and
  A~Al{\`u}.
\newblock Nonreciprocal graphene devices and antennas based on spatiotemporal
  modulation.
\newblock {\em IEEE Antennas and Wireless Propagation Letters}, 15:1529--1532,
  2015.

\bibitem{dinc2017synchronized}
Tolga Dinc, Mykhailo Tymchenko, Aravind Nagulu, Dimitrios Sounas, Andrea Alu,
  and Harish Krishnaswamy.
\newblock Synchronized conductivity modulation to realize broadband lossless
  magnetic-free non-reciprocity.
\newblock {\em Nature Communications}, 8(1):795, 2017.

\bibitem{novoselov20162d}
KS~Novoselov, A~Mishchenko, A~Carvalho, and AH~Castro Neto.
\newblock 2d materials and van der waals heterostructures.
\newblock {\em Science}, 353(6298):aac9439, 2016.

\bibitem{basov2016polaritons}
DN~Basov, MM~Fogler, and FJ~Garc{\'\i}a De~Abajo.
\newblock Polaritons in van der waals materials.
\newblock {\em Science}, 354(6309):aag1992, 2016.

\bibitem{koppens2011graphene}
Frank~HL Koppens, Darrick~E Chang, and F~Javier Garc{\'\i}a~de Abajo.
\newblock Graphene plasmonics: a platform for strong light--matter
  interactions.
\newblock {\em Nano letters}, 11(8):3370--3377, 2011.

\bibitem{PhysRevB.84.195446}
A.~Yu. Nikitin, F.~Guinea, F.~J. Garcia-Vidal, and L.~Martin-Moreno.
\newblock Fields radiated by a nanoemitter in a graphene sheet.
\newblock {\em Phys. Rev. B}, 84:195446, Nov 2011.

\bibitem{grigorenko2012graphene}
AN~Grigorenko, Marco Polini, and KS~Novoselov.
\newblock Graphene plasmonics.
\newblock {\em Nature Photonics}, 6(11):749, 2012.

\bibitem{ju2011graphene}
Long Ju, Baisong Geng, Jason Horng, Caglar Girit, Michael Martin, Zhao Hao,
  Hans~A Bechtel, Xiaogan Liang, Alex Zettl, Y~Ron Shen, et~al.
\newblock Graphene plasmonics for tunable terahertz metamaterials.
\newblock {\em Nature Nanotechnology}, 6(10):630, 2011.

\bibitem{vakil2011transformation}
Ashkan Vakil and Nader Engheta.
\newblock Transformation optics using graphene.
\newblock {\em Science}, 332(6035):1291--1294, 2011.

\bibitem{slipchenko2013analytical}
Tetiana~M Slipchenko, ML~Nesterov, Luis Martin-Moreno, and A~Yu Nikitin.
\newblock Analytical solution for the diffraction of an electromagnetic wave by
  a graphene grating.
\newblock {\em Journal of Optics}, 15(11):114008, 2013.

\bibitem{huidobro2016graphene}
Paloma~A Huidobro, Matthias Kraft, Stefan~A Maier, and John~B Pendry.
\newblock Graphene as a tunable anisotropic or isotropic plasmonic metasurface.
\newblock {\em ACS nano}, 10(5):5499--5506, 2016.

\bibitem{poumirol2017electrically}
Jean-Marie Poumirol, Peter~Q Liu, Tetiana~M Slipchenko, Alexey~Y Nikitin, Luis
  Martin-Moreno, J{\'e}r{\^o}me Faist, and Alexey~B Kuzmenko.
\newblock Electrically controlled terahertz magneto-optical phenomena in
  continuous and patterned graphene.
\newblock {\em Nature Communications}, 8:14626, 2017.

\bibitem{chen2011controlling}
Chi-Fan Chen, Cheol-Hwan Park, Bryan~W Boudouris, Jason Horng, Baisong Geng,
  Caglar Girit, Alex Zettl, Michael~F Crommie, Rachel~A Segalman, Steven~G
  Louie, et~al.
\newblock Controlling inelastic light scattering quantum pathways in graphene.
\newblock {\em Nature}, 471(7340):617, 2011.

\bibitem{efetov2010controlling}
Dmitri~K Efetov and Philip Kim.
\newblock Controlling electron-phonon interactions in graphene at ultrahigh
  carrier densities.
\newblock {\em Physical Review Letters}, 105(25):256805, 2010.

\bibitem{li2014ultrafast}
Wei Li, Bigeng Chen, Chao Meng, Wei Fang, Yao Xiao, Xiyuan Li, Zhifang Hu,
  Yingxin Xu, Limin Tong, Hongqing Wang, et~al.
\newblock Ultrafast all-optical graphene modulator.
\newblock {\em Nano letters}, 14(2):955--959, 2014.

\bibitem{tasolamprou2019experimental}
Anna~C Tasolamprou, Anastasios~D Koulouklidis, Christina Daskalaki,
  Charalampos~P Mavidis, George Kenanakis, George Deligeorgis, Zacharias
  Viskadourakis, Polina Kuzhir, Stelios Tzortzakis, Maria Kafesaki, et~al.
\newblock Experimental demonstration of ultrafast thz modulation in a
  graphene-based thin film absorber through negative photoinduced conductivity.
\newblock {\em ACS photonics}, 6(3):720--727, 2019.

\bibitem{banszerus2015ultrahigh}
Luca Banszerus, Michael Schmitz, Stephan Engels, Jan Dauber, Martin Oellers,
  Federica Haupt, Kenji Watanabe, Takashi Taniguchi, Bernd Beschoten, and
  Christoph Stampfer.
\newblock Ultrahigh-mobility graphene devices from chemical vapor deposition on
  reusable copper.
\newblock {\em Science advances}, 1(6):e1500222, 2015.

\bibitem{gonccalves2016introduction}
Paulo Andr{\'e}~Dias Gon{\c{c}}alves and Nuno~MR Peres.
\newblock {\em An introduction to graphene plasmonics}.
\newblock World Scientific, 2016.

\bibitem{iranzo2018probing}
David~Alcaraz Iranzo, S{\'e}bastien Nanot, Eduardo~JC Dias, Itai Epstein, Cheng
  Peng, Dmitri~K Efetov, Mark~B Lundeberg, Romain Parret, Johann Osmond,
  Jin-Yong Hong, et~al.
\newblock Probing the ultimate plasmon confinement limits with a van der waals
  heterostructure.
\newblock {\em Science}, 360(6386):291--295, 2018.

\bibitem{alonso2017acoustic}
Pablo Alonso-Gonz{\'a}lez, Alexey~Y Nikitin, Yuanda Gao, Achim Woessner, Mark~B
  Lundeberg, Alessandro Principi, Nicol{\`o} Forcellini, Wenjing Yan, Sa{\"u}l
  V{\'e}lez, Andreas~J Huber, et~al.
\newblock Acoustic terahertz graphene plasmons revealed by photocurrent
  nanoscopy.
\newblock {\em Nature Nanotechnology}, 12(1):31, 2017.

\bibitem{morgado2017negative}
Tiago~A Morgado and M{\'a}rio~G Silveirinha.
\newblock Negative landau damping in bilayer graphene.
\newblock {\em Physical Review Letters}, 119(13):133901, 2017.

\bibitem{PhysRevB.97.085419}
Tobias Wenger, Giovanni Viola, Jari Kinaret, Mikael Fogelstr\"om, and Philippe
  Tassin.
\newblock Current-controlled light scattering and asymmetric plasmon
  propagation in graphene.
\newblock {\em Phys. Rev. B}, 97:085419, Feb 2018.

\bibitem{ghafarian2016millimetre}
Naimeh Ghafarian, Hamed Majedi, and Safieddin Safavi-Naeini.
\newblock Millimetre-wave and terahertz amplification in a travelling wave
  graphene structure.
\newblock {\em IEEE Journal of Selected Topics in Quantum Electronics},
  23(1):179--187, 2016.

\bibitem{wilson_low2018}
Josh Wilson, Fadil Santosa, Misun Min, and Tony Low.
\newblock Temporal control of graphene plasmons.
\newblock {\em Phys. Rev. B}, 98:081411, Aug 2018.

\bibitem{sun2016adiabatic}
Zhiyuan Sun, DN~Basov, and MM~Fogler.
\newblock Adiabatic amplification of plasmons and demons in 2d systems.
\newblock {\em Physical Review Letters}, 117(7):076805, 2016.

\bibitem{kaminer2016efficient}
Ido Kaminer, Yaniv~Tenenbaum Katan, Hrvoje Buljan, Yichen Shen, Ognjen Ilic,
  Josu{\'e}~J L{\'o}pez, Liang~Jie Wong, John~D Joannopoulos, and Marin
  Solja{\v{c}}i{\'c}.
\newblock Efficient plasmonic emission by the quantum {\v{c}}erenkov effect
  from hot carriers in graphene.
\newblock {\em Nature Communications}, 7:11880, 2016.

\bibitem{knight1992optical}
PL~Knight and A~Miller.
\newblock {\em Optical solitons: theory and experiment}, volume~10.
\newblock Cambridge University Press, 1992.

\bibitem{cassedy1965waves}
Edward~S Cassedy.
\newblock Waves guided by a boundary with time—space periodic modulation.
\newblock In {\em Proceedings of the Institution of Electrical Engineers},
  volume 112, pages 269--279. IET, 1965.

\bibitem{stauber2007electronic}
T~Stauber, NMR Peres, and F~Guinea.
\newblock Electronic transport in graphene: A semiclassical approach including
  midgap states.
\newblock {\em Physical Review B}, 76(20):205423, 2007.

\end{thebibliography}

\end{document}